\documentclass[aps,showpacs,twocolumn]{revtex4}
\usepackage{epsfig}

\begin{document}

\title{The effect of hidden color channels on Nucleon-Nucleon
interaction\footnote{Email address: fgwang@chenwang.nju.edu.cn}}

\author{Fan Wang$^a$, Jialun Ping$^b$, Hongxia Huang$^b$}

\affiliation{$^a$J-CPNPC (Joint Center for Particle Nuclear Physics
and Cosmology, Nanjing University and Purple Mountain Observatory,
Chinese Academy of Sciences) Nanjing 210093, P.R. China}

\affiliation{$^b$Department of Physics, Nanjing Normal University,
Nanjing 210097, P.R. China}

\begin{abstract}
This letter reports the nucleon-nucleon($NN$) interaction obtained
from multi-channel, including hidden color channels, coupling quark
model calculation. The results show that the hidden color channels
coupling provides the intermediate range attraction which is usually
assumed to be due to multi-$\pi$ or $\sigma$ meson exchange and that
the short and intermediate range $NN$ interaction can be described
solely by the fundamental quark-gluon degree of freedom of QCD.
\end{abstract}



\maketitle

The $NN$ interaction has been studied for more than 70 years since
the discovery of neutron and the meson exchange model proposed by H.
Yukawa. The effective one boson exchange model (OBE)~\cite{mah,des}
and the chiral perturbation theory (ChPT)~\cite{mei} fit the $NN$
interaction data below the $\pi$ production threshold quantitatively
well but with $\sim$30 (ChPT) to more than 40 (OBE) adjustable
parameters. Quark models~\cite{sal,QDCSM} fit the $NN$ interaction
data but quantitatively not as well as OBE and ChPT. However they
use much less adjustable parameters, 9 for~\cite{QDCSM}, and fit the
hadron spectroscopy simultaneously. Lattice QCD has got the static
$qq$ interaction of 2, 3, 4 and 5 quark systems and the qualitative
$NN$ interaction in the quenched approximation recently~\cite{lat}.
The static interaction can be expressed as color Coulomb plus the
linear confinement proportional to the minimum length of the color
flux with junctions. Unquenched lattice results only modify the
linear confinement with color screening but have not found the trace
of meson exchange yet~\cite{lat}. One expects lattice QCD will
describe $NN$ interaction by means of the fundamental quark-gluon
degree of freedom of QCD directly.

The short and even the intermediate range $NN$ interaction should be
related to the nucleon internal structure as emphasized by P.W.
Anderson~\cite{and}. The long standing fact that the $^3S_1$ and
$^1S_0$ $NN$ interactions are similar to the hydrogen molecular
forces in the spin singlet and triplet states respectively except
the energy and length scale difference, which certainly calls for an
explanation but has never been explained in OBE and ChPT. Quark
model should be able to do this job, because the $NN$ interaction is
treated as a remnant of the fundamental color interaction due to
gluon exchange between color quarks of two color singlet nucleons,
which is quite the same as the hydrogen molecular force is a remnant
of the fundamental electro-magnetic interaction due to photon
exchange between charged particles of two electric neutral hydrogen
atoms. However it is well known that the effective one gluon
exchange Breit-Fermi interaction plus two body confinement quark
models only obtained the $NN$ short range repulsion but can not
describe the intermediate and long range attraction. Chiral quark
model~\cite{sal} introduced the $\pi$ and $\sigma$ meson exchange to
provide the intermediate and long range attraction. We proposed a
model~\cite{QDCSM} in which we introduced the quark mutual
delocalization (or percolation) to describe the spacial mutual
distortion of interacting nucleons and a phenomenological color
screening to model the color space mutual distortion (hidden color
channel effect). In this report we show the $NN$ scattering results
obtained from the multi channel coupling calculation with hidden
color channels coupling directly. These results show that the hidden
color channels coupling does give rise to the $NN$ intermediate
range attraction and that the short and intermediate range $NN$
interaction can be described by the fundamental quark-gluon degree
of freedom directly. The mechanism of the short and intermediate
range $NN$ interaction is quite the same as that of the hydrogen
molecular force and so provide a natural explanation of the
similarity between nuclear and molecular forces.

The $NN$ system (a six valence quark system) can be consisted of two
color singlet nucleons as in the usual hadron degree of freedom
description, but also of two color octet nucleons coupled to an
overall color singlet six quark state as shown in Fig. 1. The latter
is called hidden color channel and because of color confinement,
these hidden color channels exist in the two nucleon overlap region
only. The quark cluster model channel wave function can be expressed
as,

\begin{eqnarray}
\Psi(NN) &=&
\mathcal{A}\left[[\psi(N_1)\psi(N_2)]_{CIS}F(\mathbf{R})\right],
\nonumber \\
\psi(N_i) &=& \chi_{c_i}\zeta_{I_iS_i}\phi(\mbox{\boldmath$\xi$}_i),
\end{eqnarray}

where $\mathcal{A}$ is the anti-symmetrization operator, $\psi(N_i)$
is the nucleon internal wave function, $\chi_{c_i}$ is the color
part which can be color singlet or octet, $\zeta_{I_iS_i}$ is the
$SU(4)\supset SU^{\tau}(2)\times SU^{\sigma}(2)$ spin-isospin part,
$\phi(\mbox{\boldmath$\xi$}_i)$ is the orbital part and
$\mbox{\boldmath$\xi$}_i$ is the internal Jacobian coordinates. In
order to simplify numerical calculation we assume it is a production
of Gaussian function with a size parameter $b$.
$[\cdot\cdot\cdot]_{CIS}$ means coupling the individual nucleon
color-isospin-spin into overall color singlet, total isospin $I$ and
spin $S$ with $SU^c(3),SU^{\tau}(2),SU^{\sigma}(2)$ Clebsch-Gordan
Coefficients. $F(\mathbf{R})$ is the relative orbital wave function
and $\mathbf{R}$ is the relative coordinate between two nucleon
center of mass coordinates,
$\mathbf{R}=\mathbf{R}_{N_1}-\mathbf{R}_{N_2}$. The quark
delocalization description can be done as the same in~\cite{QDCSM}
and we will not repeat here to save the space.

\begin{center}
\epsfxsize=3.2in \epsfbox{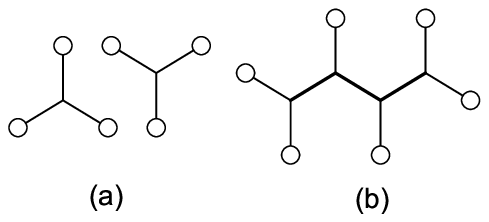}

Fig. 1. color structures of $NN$ system.
\end{center}

The model Hamiltonian of the $NN$ system is chosen to be,
\begin{eqnarray}
&& H = \sum_{i=1}^6 \left(m_i+\frac{\mathbf{p}_i^2}{2m_i}\right)
-T_c
+\sum_{i<j} \left( V^{G}_{ij}+V^{\pi}_{ij}+V^{C}_{ij}\right), \nonumber \\
&& V^{G}_{ij}= \frac{\alpha_s}{4} \mbox{\boldmath$\lambda$}_i \cdot
\mbox{\boldmath$\lambda$}_j
\left[\frac{1}{r_{ij}}-\frac{\pi}{m_q^2}\left(1+\frac{2}{3}
\mbox{\boldmath$\sigma$}_i\cdot \mbox{\boldmath$\sigma$}_j \right)
\delta(\mathbf{r}_{ij}) \right. \nonumber
\\
&& ~~~~~~~~~~~~~~~~~- \left.
\frac{3}{4m_q^2r^3_{ij}}S_{ij}+v^{G,LS}_{ij}\right],
\nonumber \\
&& v^{G,LS}_{ij} = -\frac{1}{8m_q^2}\frac{3}{r_{ij}^3}[{\mathbf
r}_{ij} \times ({\mathbf p}_i-{\mathbf p}_j)] \cdot(\mbox{\boldmath$
\sigma$}_i+\mbox{\boldmath$\sigma$}_j),
 \\
&& V^{\pi}_{ij}= \frac{\alpha_{ch}}{3}
\frac{\Lambda^2}{\Lambda^2-m_{\pi}^2}m_\pi \left\{ \left[ Y(m_\pi
r_{ij})- \frac{\Lambda^3}{m_{\pi}^3}Y(\Lambda r_{ij}) \right]
\right.
\nonumber \\
&& ~~~~~~\left. \mbox{\boldmath$\sigma$}_i \cdot
\mbox{\boldmath$\sigma$}_j+\left[ H(m_\pi
r_{ij})-\frac{\Lambda^3}{m_\pi^3} H(\Lambda r_{ij})\right] S_{ij}
\right\} \mbox{\boldmath$\tau$}_i
\cdot \mbox{\boldmath$\tau$}_j,  \nonumber \\
&& V^{C}_{ij}= -a_{c}\mbox{\boldmath$\lambda$}_{i} \cdot
\mbox{\boldmath$\lambda$}_{j}(r^2_{ij}+V_0), \nonumber
\\
&& S_{ij}  =  \frac{(\mbox{\boldmath$\sigma$}_i \cdot {\mathbf
r}_{ij}) (\mbox{\boldmath$\sigma$}_j \cdot {\mathbf
r}_{ij})}{r_{ij}^2}-\frac{1}{3}~\mbox{\boldmath$\sigma$}_i \cdot
\mbox{\boldmath$\sigma$}_j. \nonumber
\end{eqnarray}
Here $S_{ij}$ is quark tensor operator, $Y(x)$ and $H(x)$ are
standard Yukawa functions, $T_c$ is the kinetic energy of the center
of mass. All other symbols have their usual meaning.

The model parameters are fixed as follows: the $u,d$-quark mass
difference is neglected and $m_u$=$m_d$ is assumed to be exactly
$1/3$ of the nucleon mass, namely $m_u$=$m_d$=$313~$MeV, the $\pi$
mass takes the experimental value, the $\Lambda_{\pi}$ takes the
same values as in Ref.\cite{sal}, namely $\Lambda_{\pi}$=4.2
fm$^{-1}$. The chiral coupling constant $\alpha_{ch}$ is determined
from the $\pi NN$ coupling constant as usual. The rest parameters
$a_c$, $V_0$, and $\alpha_s$ are determined by fitting the nucleon
and $\Delta$ masses and the stability of nucleon mass with respect
to the variation of the nucleon size $b$. All parameters used are
listed in Table 1.

\begin{center}
Table 1. Quark model parameters.

\nopagebreak
\begin{tabular}{ccc}\hline
quark masses & $m_{u,d}$ (MeV) & 313 \\
\hline
nucleon size & $b$ (fm)   & 0.518 \\
\hline
             & $m_{\pi}$ (fm$^{-1}$) & 0.71 \\
  $\pi$      & $\Lambda_{\pi}$ (fm$^{-1}$) & 4.2 \\
             & $\alpha_{ch}$  & 0.027 \\
\hline
 Confinement & $a_c$(MeV fm$^{-2}$) &56.75 \\
             & $V_0$ (fm$^2$)       &-1.3590 \\
\hline
    OGE      & $\alpha_s$     & 0.485 \\
\hline
\end{tabular}
\end{center}

This study is focused on the hidden color channel effect and so in
order to simplify the numerical calculation we approximate the
nucleon internal orbital wave function by a product of two Gaussian
functions of the Jacobian coordinates. This approximation will make
the exchange interaction matrix elements decreasing with R as
$e^{-R^2/2b^2}$ and so definitely can not describe the one $\pi$
exchange Yukawa interaction. Therefore we include the one $\pi$
exchange in our model Hamiltonian. The many body linear confinement
interaction as obtained in lattice QCD is also approximated by two
body quadratic one as shown in Eq.(2) and the color screening due to
$q\bar{q}$ excitation observed in unquenched lattice QCD is
neglected too. These approximations will hinder the quark mutual
delocalization and quite possible reduce the effective $NN$
attraction as otherwise due to quark delocalization. All of these
shortcomings are left for further improvement.

The Kamimura variational method~\cite{kam} is used to do multi
channel scattering calculation. The channels included in different
partial waves are listed in Table 2. The calculated partial wave
phase shifts and the SP07~\cite{SP07} data points are shown in
Fig.2-5. The direct extension of the color dependent two body
confinement model can not describe the $NN$ scattering
quantitatively well even after including hidden color channels
coupling as can be seen from the first set results (the solid lines
in Figs.2-5). The color dependent two body confinement interaction
is consistent with the lattice QCD results only for two and three
quark systems in color singlet states but inconsistent with the many
body interaction obtained in lattice QCD ones for multi-quark
systems \cite{lat}. For multi-quark systems and color octet
nucleons, quark pairs are not always in color antisymmetric state
but also color symmetric ones. The color factor
{\boldmath$\lambda$}$_{i} \cdot$ {\boldmath$\lambda$}$_{j}$ will
give rise to anti-confinement interaction for symmetric quark pairs
\cite{deconfine}. There is no sound theoretical reason to extend the
color dependent two body confinement interaction to multi-quark
system. We adjust the confinement interaction strength $a_c$ for
every partial waves to fit the $NN$ phase shifts of SP07 by two
recipes. recipe 1: adjusting the $a_c$ to $ka_c$ for the coupling
between hidden color channels and the color singlet channels only;
recipe 2: adjusting the confinement strength $a_c$ to $ka_c$ not
only for the coupling between hidden color channels and the color
singlet ones but also for the hidden color channel themselves. The
best fit of the multiplicative factor $k$'s are listed in Table 2
too. One can see from the dotted (recipe 1) and dashed (recipe 2)
lines in Figs.2-5 that by including the hidden color channels and
adjusting the color confinement interaction strength $a_c$, both
adjusting recipes can fit the $NN$ scattering phase shifts
quantitatively. We take these results as an indication that the
short and intermediate range $NN$ interaction can be described
solely by the fundamental quark-gluon degree of freedom. It is the
nucleon internal structure and its distortion both in orbital and
color spaces which give rise to the $NN$ short range repulsion and
intermediate range attraction and these are quite the same as the
atomic internal structure and its distortion in orbital and electric
charge spaces which give rise to the hydrogen molecular interaction.

\begin{widetext}
\begin{center}
Table 2. The channels used in $NN$ scattering calculations and the
confinement strength for each channel.

\begin{tabular}{c|lcc}
\hline
 $~I~~J~$ & \mbox{~~~~~~~}\hspace{1in} channels & ~recipe~ & ~~~$k$~~~   \\ \hline
 ~0~~1~ & $^3S_1(^3D_1): NN,\Delta\Delta, ~^2\Delta_8~ ^2\Delta_8,
 ^4N_8 ~^4N_8,~^4N_8 ~^2N_8,~^2N_8 ~^2N_8
 $ & 1 & $1.40$  \\\cline{3-4}
 & $^7D_1: \Delta\Delta,~^4N_8~^4N_8$ & 2 & $1.38$  \\ \cline{2-4}
& $^1P_1:
NN,\Delta\Delta,~^2\Delta_8~^2\Delta_8,~^4N_8~^4N_8,~^2N_8~^2N_8$ &
1 & $1.80$  \\\cline{3-4}
 & $^5P_1: \Delta\Delta,~^4N_8 ~^4N_8, ~^4N_8~^2N_8$ & 2 & $1.70$  \\  \hline
~0~~2~ & $^3D_2:
NN,\Delta\Delta,~^2\Delta_8~^2\Delta_8,~^4N_8~^4N_8,~^4N_8~^2N_8,~^2N_8~^2N_8$
& 1 & $1.00$  \\\cline{3-4}
 & $^7D_2: \Delta\Delta,~^4N_8~^4N_8$ & 2 & $1.00$  \\ \hline
 ~0~~3~ & $^3D_3:
NN,\Delta\Delta,~^2\Delta_8~^2\Delta_8,~^4N_8~^4N_8,~^4N_8~^2N_8,~^2N_8~^2N_8$
& 1 & $2.40$  \\\cline{3-4}
 & $^7S_3(^7D_3): \Delta\Delta,~^4N_8~^4N_8$ & 2 & $2.20$  \\ \hline
~1~~0~ & $^1S_0:
NN,\Delta\Delta,~^2\Delta_8~^2\Delta_8,~^4N_8~^4N_8,~^2N_8~^2\Delta_8,~^2N_8~^2N_8$
& 1 & $1.42$  \\\cline{3-4}
 & $^5D_0: N\Delta,\Delta\Delta,~^4N_8~^2\Delta_8,~^4N_8~^4N_8,~^4N_8~^2N_8$ & 2 & $1.39$  \\ \cline{2-4}
& $^3P_0:
NN,N\Delta,\Delta\Delta,~^2\Delta_8~^2\Delta_8,~^4N_8~^2\Delta_8,~^4N_8~^4N_8,~^4N_8~^2N_8,$
& 1 & $1.10$  \\\cline{3-4}
 & ~~~~~~~$^2N_8~^2\Delta_8,~^2N_8~^2N_8$ & 2 & $1.10$  \\  \hline
~1~~1~ & $^3P_1:
NN,N\Delta,\Delta\Delta,~^2\Delta_8~^2\Delta_8,~^4N_8~^2\Delta_8,~^4N_8~^4N_8,~^4N_8~^2N_8,$
& 1 & $1.35$  \\\cline{3-4}
 & ~~~~~~~$^2N_8~^2\Delta_8,~^2N_8~^2N_8$ & 2 & $1.28$  \\  \hline
~1~~2~ & $^1D_2:
NN,\Delta\Delta,~^2\Delta_8~^2\Delta_8,~^4N_8~^4N_8,~^2N_8~^2\Delta_8,~^2N_8~^2N_8$
& 1 & $2.00$  \\\cline{3-4}
 & $^5S_2(^5D_2): N\Delta,\Delta\Delta,~^4N_8~^2\Delta_8,~^4N_8~^4N_8,~^4N_8~^2N_8$ & 2 & $1.85$  \\ \cline{2-4}
& $^3P_2:
NN,N\Delta,\Delta\Delta,~^2\Delta_8~^2\Delta_8,~^4N_8~^2\Delta_8,~^4N_8~^4N_8,~^4N_8~^2N_8,$
& 1 & $1.75$  \\\cline{3-4}
 & ~~~~~~~$^2N_8~^2\Delta_8,~^2N_8~^2N_8$ & 2 & $1.66$  \\  \hline
\end{tabular}
\end{center}
\end{widetext}

\begin{center}
\epsfxsize=3.0in \epsfbox{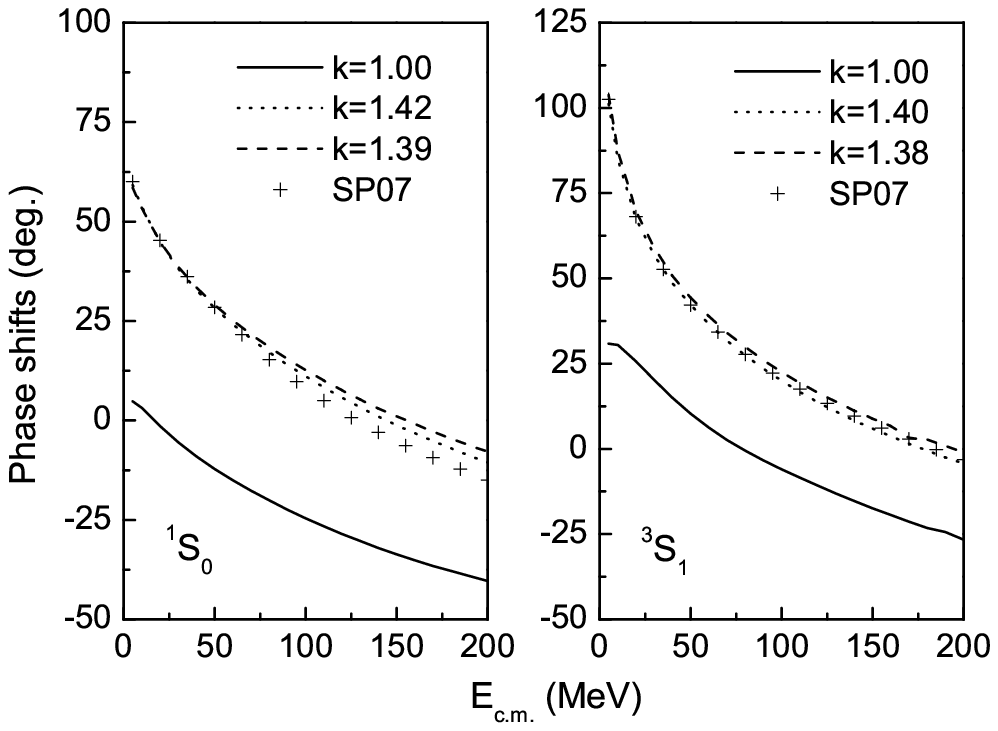}

Fig.2. The $S$-wave phase shifts of $NN$ scattering.
\end{center}

The model is also used to study the properties of deuteron. The
results are given in Table 3. The binding energy can be reproduced
(we didn't fine tune the strength of color confinement to get the
better binding energy) as well as the $D$-wave component in the
deuteron. However the root mean square radius is too small compared
to experimental value. This shortcoming may be due to the fact that
the deuteron wave function is only extending to 10 fm in this many
channels coupling calculation.
\begin{center}
\epsfxsize=3.0in \epsfbox{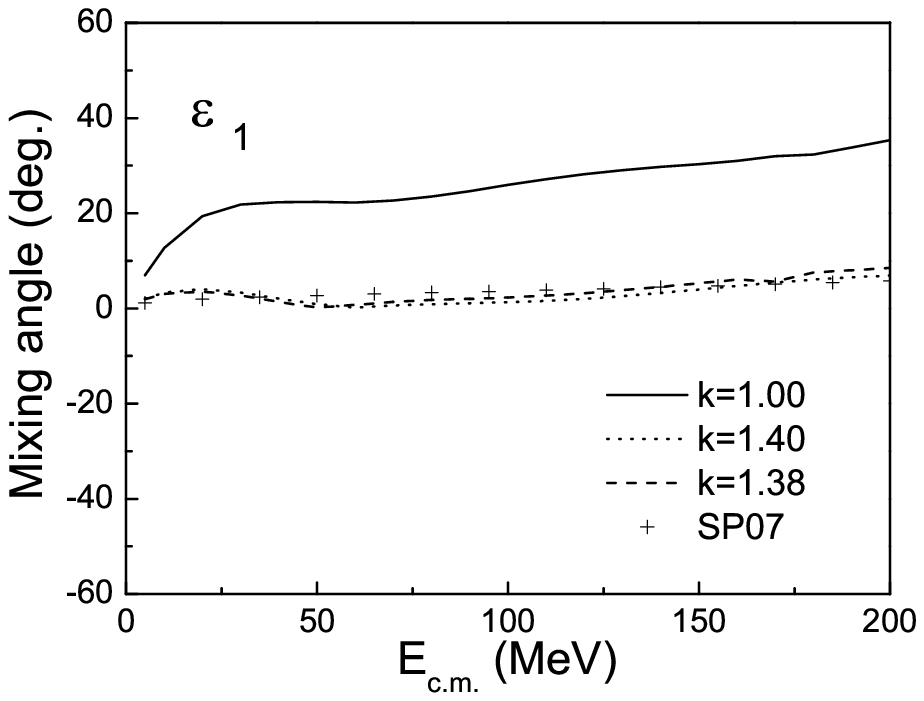}

Fig.3. The mixing angles $\epsilon_1$ of $NN$ scattering.

\end{center}

\begin{center}
Table 3. The properties of deuteron.

\nopagebreak

\begin{tabular}{ccccc} \\ \hline
recipes  & ~$a_c$(MeV fm$^{-2}$)~ & ~$E_B$(MeV)~ & ~rms(fm)~ &
~$P_D$
\\
\hline 1 & 79.45 & 1.0 & 1.2 & 4\% \\ \hline 2 & 78.315 & 2.2 & 1.1
& 4\%
\\ \hline
\end{tabular}
\end{center}

\begin{widetext}
\begin{center}
\epsfxsize=4.9in \epsfbox{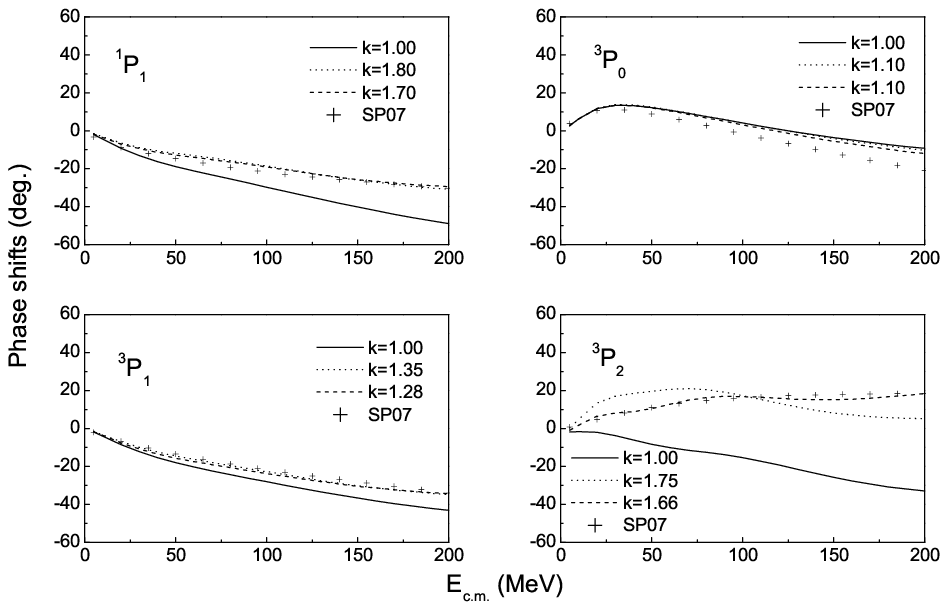}

Fig.4. The $P$-wave phase shifts of $NN$ scattering.

\epsfxsize=4.9in \epsfbox{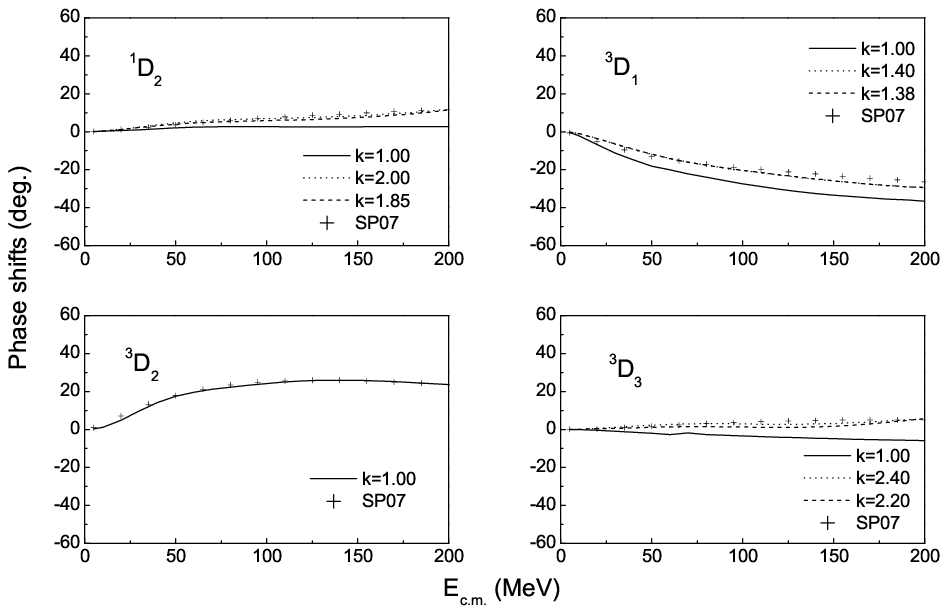}

Fig. 5. The $D$-wave phase shifts of $NN$ scattering.
\end{center}
\end{widetext}

There are problems remained and need to be studied further: (i) Up
to now we don't know the confinement interaction between different
color structures, this quark model study shows one need a
phenomenological color confinement interaction different for
different color partial waves but this might be artificial due to
those approximations used in this calculation. (ii) The color
screening effect due to unquench or $q\bar{q}$ excitation should be
studied which might improve the above results. The effect of
multi-body confinement interaction as obtained in the lattice QCD
calculation should be studied too, because there are indications
that it will effect the multi quark systems~\cite{deng}. (iii) The
$\pi$ exchange is directly included in the Hamiltonian, is it
possible to describe its effect directly from quark-gluon degree of
freedom, as lattice QCD did~\cite{lat}, by means of a more realistic
nucleon internal orbital wave function and a more sophisticated
color confinement interaction as mentioned above. Such a calculation
is quite involved numerically but it seems to be worth to devote.

This work is supported in part by NSFC grant 90503011, 10435080,
10375030, 10775072 and the Research Fund for the Doctoral Program of
Higher Education of China under Grant No. 20070319007.

\end{document}